\documentclass[pre,twocolumn]{revtex4}

\usepackage[dvips]{graphicx}
\usepackage{amssymb}
\usepackage{amsmath}

\begin{document}

\title{A surface force apparatus for nanorheology under large shear strain}

\author{Lionel Bureau}
 \email{bureau@insp.jussieu.fr}
\affiliation{Institut des Nanosciences de Paris, UMR 7588 CNRS-Universit\'e Paris 6, 140 rue de Lourmel, 75015 Paris, France}


\begin{abstract}
We describe a surface force apparatus designed to probe the rheology of a nanoconfined medium under large shear amplitudes (up to 500 $\mu$m).  The instrument can be
operated in closed-loop, controlling either the applied normal load or the thickness of the medium during shear experiments. Feedback control allows to greatly extend 
the range of confinement/shear strain attainable with the surface force apparatus. The performances of the instrument are illustrated using hexadecane as the confined medium.
\end{abstract}

\maketitle

\section{Introduction}
\label{sec:intro}

The surface force apparatus (SFA) has been developed more than thirty years ago to probe surface interactions through a
direct determination of force {\it vs} separation between two atomically smooth surfaces \cite{TW,IT}.
This technique, originally designed to measure forces between surfaces separated by an air gap, has then been
extended to force measurements across liquids of various nature \cite{IA,HI}. It appears, from the large body
of work done in the field,  that many liquids, when 
confined down to molecular thickness, tend to order into layers parallel to the 
confining walls \cite{Ibook,HI2,HKC1,KK1,ZG}. 
  
The existence of such a molecular layering of simple liquids is at the origin of a number of studies which 
addressed the question of the lubricating 
properties of nanoconfined fluids subjected to shear. To do so, different versions of the SFA have been
designed, that allow for tangential displacement of one of the confining surfaces and for shear force measurement. The 
studies performed with these friction devices focus either on the viscoelastic behavior of the confined fluid ({\it 
i.e.} on the response to small amplitude oscillatory shear) \cite{RDG,granick2} or on its frictional response when one of the surfaces is
driven at constant speed in one direction \cite{IMH1,KK2}. The following central result emerges: when confined to thicknesses on the order
of five molecular diameters, many liquids 
exhibit a finite yield stress and strongly non-newtonian flow properties. Moreover, such a 
solidlike behavior is most sensitive to the number
of molecular layers between the walls \cite{KK2,IMH2}. Maintaining a constant thickness, to within a few angstroms, of the medium during shear is
thus of primary importance in these experiments. 

Now, unavoidable mechanical imperfections of the instruments usually
cause a slight non parallel motion of the surfaces, which typically results in thickness variations of 10--30 \AA~for a shear amplitude of 1 $\mu$m
\cite{tirrell1,perez}. For a weakly confined medium, of thickness $h\sim 100$ \AA,  the useful range of shear amplitude
is thus limited to about 500 nm if one prescribes relative variations of $h$ of less than a few percents. 
For a medium confined under a finite normal force applied through a spring of typical stiffness 300 N.m$^{-1}$, 
if one wants the above-mentioned non-parallelism to cause an ``acceptable '' load variation of, say, 5\%, over a shear amplitude of 
100 $\mu$m, the applied load must therefore be of at least 1 mN. Such a load level corresponds to strongly confined regimes 
where the thickness is typically 1--2 molecular layers.

Such conditions are not limiting as long as molecules of simple structure are confined, for which a shear amplitude of about 1 $\mu$m is 
enough to establish a steady-state regime. However, they may become more of a concern when investigating the rheological properties 
of more complex molecules like branched hydrocarbons or polymers, which exhibit transients corresponding to sliding distances of more 
than 100 $\mu$m \cite{drummond,gourdon,luengo}.
     
To our knowledge, only two devices have been designed to allow for compensation of non-parallel motion and thus extend the
range of shear amplitude and confinement to be investigated \cite{KK1,KK2,perez}. In the present paper, we describe a surface force apparatus with 
unique performances in terms of intersurface distance stability over large shear amplitudes. This apparatus relies on a
well-established mechanical design \cite{HKC2}, and uses multiple beam interferometry to determine the thickness of the confined medium.
The particularity of the SFA presented here is to allow for time-resolved distance and normal force measurements that are
used as input signals of a digital feedback loop. This enables us to operate the apparatus either at constant thickness or constant force 
during shear motion over up to 500 $\mu$m.
    
\section{General description}
\label{sec:general}

\begin{figure}[htbp]
$$
\includegraphics[width=7cm]{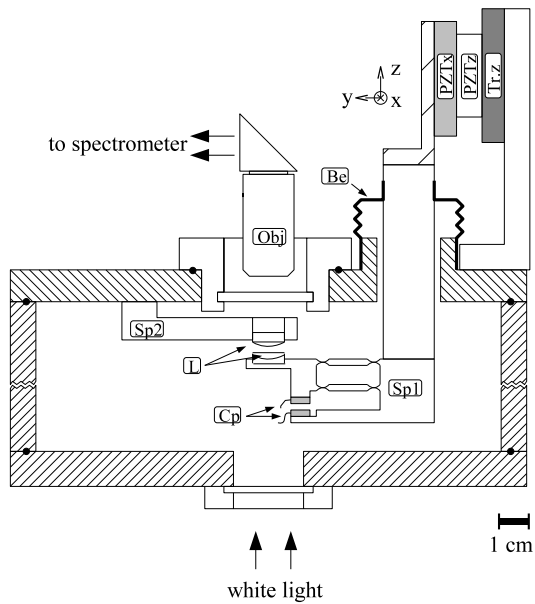}
$$
\caption{Mechanical arrangement of the surface force apparatus. Two crossed cylindrical lenses (L) are mounted in springs (Sp1 and Sp2), the deflexion 
of which is measured by means of capacitive displacement sensors (Cp). The normal load spring (Sp1)  is attached to a shaft that can be moved horizontally by means of a 
long travel piezoactuator (PZTx), and vertically $via$ a two stage mechanism composed of a piezoactuator (PZTz) and a motorized translation stage (Tr.z). A rubber bellow (Be) is used
to ensure air-tightness of the vessel while allowing for normal and tangential displacement of the shaft. }
\label{fig:fig1}
\end{figure}

The general design of the surface force apparatus is given on Fig. \ref{fig:fig1}. The medium of interest is 
confined between two molecularly smooth mica sheets
 glued on crossed cylindrical lenses, of radius of curvature 1 cm.
The mechanical arrangement is very similar to that proposed by Parker and Christenson \cite{HKC2}: the SFA is made of an aluminium alloy
cylindrical
cell closed by a horizontal circular flange on which all the mechanical parts are mounted. 

The lower mica surface is held at one end of a 
vertical 
spring, the other end of which is attached to a shaft that can be moved vertically ($z$ axis) and horizontally ($x$ axis). Coarse vertical
motion is produced by a miniature translation stage equipped with a stepper motor (Physik Instruments PI-M111.12s, total travel 15 mm,
positioning accuracy 100 nm with controller PI-C630). Fine vertical approach of the surfaces is made by means of a hysteresis-free
piezoactuator (PI-P753.11C and control electronics PI-E509.C3A ) of 12 $\mu$m range, which includes a built-in capacitive sensor
 allowing for 
closed-loop operation with a positioning accuracy of 0.5 \AA. Motion along the $x$ axis is produced by a piezoactuator using the 
same technology  but with a total travel of 500 $\mu$m (PI-P625.1CD and controller PI-E509.C1A). 

The upper mica surface is held by a horizontal spring which serves as a shear force sensor. 

The apparatus is placed on an active anti-vibration table (MOD-1M, Halcyonics GmbH, Germany), and the whole setup is housed in a
home-made thermally regulated enclosure.  Four low-noise ventilators are used to maintain forced convection inside the enclosure, 
where the temperature can be adjusted at $\pm 0.02^{\circ}$C in the range 10--45$^{\circ}$C by means of a thermoelectric 
cooling/heating assembly controlled by a commercial PID unit (SuperCool PR-59). 

\subsection{Force measurement}

Both vertical and horizontal springs
are deformable parallelograms with flexure hinges, each of them being precision-machined from one block of highly elastic 
aluminium
alloy. Their respective stiffness  are $K_{N}=31000\pm 800$ N.m$^{-1}$ and $K_{T}=28000\pm700$
N.m$^{-1}$, and their deflections are measured by means
of capacitive displacement sensors with a 15 $\mu$m range (Physik Instruments D-015 with electronics E509.C3A) and a resolution
of 1.5 pm/$\sqrt{\text{Hz}}$. This translates into a nominal force sensitivity, in both $z$ and $x$ directions, of 
$\sim$0.05~$\mu$N/$\sqrt{\text{Hz}}$ over a range of about 450 mN. 
In practice, the force sensitivity is limited by residual vibrations of the springs and is of $\pm$2 $\mu$N on a 100 Hz bandwidth.
The maximum frequency at which force measurements
can be performed is limited by the bandwidth of the electronics, namely 3kHz.

Note that in contrast with many surface force apparatuses, in which the normal load is deduced from the
difference in the imposed displacement of the spring remote point and the displacement actually measured between the
 surfaces, our solution allows to perform force measurements independently of the distance determination.

\subsection{Distance measurement}

Intersurface distance measurement relies on multiple beam interferometry. 

The Fabry-Perot cavity formed by the silvered
mica sheets and the confined medium is shone with a collimated beam of white light produced by a 250W tungsten-halogen source (Lot-Oriel). 
Infrared radiation is filtered using a series of cold mirrors and water filters. 

The region surrounding the point of closest distance between the cylindrical lenses is imaged on the entrance slit of a spectrograph using a long working distance microscope objective 
(VLD-MPA10, numerical aperture 0.23. Nachet, France, ). The spatial resolution is diffraction limited and is of $\sim$1.5 $\mu$m.

The spectral position of the fringes of equal chromatic order (FECO) is determined with a half-meter imaging spectrometer
(Acton spectra-pro 2500i) equipped with 3 gratings of 300, 600 and 1200 g/mm. A 16-bits 2D charge coupled device camera (PIXIS 400, Princeton Instruments, 1340$\times$ 400 pixels) is placed at the exit port of the spectrometer.
With the 1200 g/mm grating, the spectral resolution is of 30 pm/pixel over a wavelength range of 40.2 nm. Good
quality images of the spectrograph exit field are obtained with exposure times $t_{\text{exp}}\geq 20$~ms. The readout 
time of the full CCD array is of 280 ms, which yields a rate of about 3 frames per second for acquisition of the FECO
along a spatial coordinate. If measurements of the FECO are made only at the point of closest distance, the readout time is that of a single row of pixel and drops to 17 ms, which, added to $t_{\text{exp}}\simeq 20$~ms, yields a maximum rate
of about 27 Hz for spectrum acquisition at a single location.

The distance between the mica surfaces is deduced from the FECO using a Labview-based implementation of fast spectral correlation (FSC) and multilayer matrix method (MMM), as 
introduced recently by Heuberger \cite{Heuberg1}. We briefly describe the protocol followed for distance measurements:

{\it (i)} the mica surfaces are brought into contact, a spectrum is acquired in the flattened contact zone, and the wavelengths $\lambda_{i}$ of the transmission maxima are determined.

{\it (ii)} for the set of $\lambda_{i}$ determined at mica-mica contact, we calculate, using the MMM, the transmissivity of the silver/mica/mica/silver layered medium for a range of plausible
mica thickness $d_{\text{mica}}$ (typically 10000--100000~\AA, with 1~\AA~steps). We choose for $d_{\text{mica}}$ the value that maximizes the function
$T=\sum_{i}T_{i}$, where $T_{i}$ is the transmissivity calculated at the wavelengths $\lambda_{i}$ \cite{Heuberg1}. We finally calculate for this value of $d_{\text{mica}}$ the transmissivity over the full experimental
wavelength window in order to check if both the number and the position of the calculated transmission maxima are in agreement with the experimental spectrum. The time needed for the determination 
of $d_{\text{mica}}$ is of a few seconds and is limited by the time for transmissivity calculation over the chosen thickness range.

{\it (iii)} the surfaces are separated by $\sim 1$ mm, and  the liquid is injected between them using a microsyringe.

{\it (iv)} the mica sheets are approached down to a separation of a few microns.

{\it (v)} the initial thickness of the liquid, $d_{0}$, is determined as in step {\it (ii)} above: we acquire a spectrum at the point of 
closest distance between the curved surfaces, locate the position of the transmission maxima, and calculate for these positions the transmissivity of the silver/mica/liquid/mica/silver 
multilayer for a range of plausible thickness (typically 0--50000~\AA, with 10~\AA~steps), assuming that the refractive index of the liquid is the bulk one. Once again, the time needed to complete this step is 
limited by the transmissivity calculation
over the range chosen for the trial values of  $d_{0}$.

{\it (vi)} once $d_{0}$ is known, at each time $t_{i}$ a new spectrum is acquired, the intersurface distance $d_{i}$ is calculated as in step (v), with a range of plausible distance 
$d_{i-1} \pm \Delta d$, where $d_{i-1}$ is the distance determined at time $t_{i-1}$. The time for transmissivity calculations using $\Delta d=100$ \AA~and 1\AA~increments is on
the order of 1 ms, and distance measurements can thus be performed at a rate limited only by spectra acquisition. 

\subsection{Data acquisition and feedback operation}

Spectra are transferred to the host computer via its universal serial bus and treated immediatly to deduce the intersurface distance as described above. The signals corresponding to the normal and tangential forces and
to the positions of the normal and tangential piezoactuators are measured by four digital multimeters (DMM 34401A, Agilent, 6$^{1/2}$ digits). The synchronization output of the PIXIS camera is used to trigger the multimeters each time a 
spectrum has been
acquired. Measurements from the DMMs are transferred to the host computer between each trigger event, {\it via} a GPIB-PCI card. 
Feedback operation can be performed using either the intersurface distance or the normal force signal as the input of a digital loop with proportional and integral control, which acts on the voltage of 
the normal piezoactuator. The gains of the loop were set as described in \cite{JB}: we first increase the proportional gain until oscillations are observed, lower it by a few percents from this value, then 
increase the integral gain in order to reduce the offset to setpoint.

\section{performances}
\label{sec:perf}

\subsection{Sample preparation}

Mica sheets are cleaved down to a thickness of 2--5 $\mu$m and cut into approximately 1 cm$^{2}$ squares by means of
surgical scissors. This samples are put on a clean mica backing substrate with one angle lying on a thin teflon ribbon 
in order to allow for subsequent easy de-adhesion. The back side of the mica samples is evaporated with a 45nm-thick
silver layer. The results presented hereafter have been obtained using glucose to glue the samples on the cylindrical lenses. 

We demonstrate the performances of the apparatus using a linear alkane as a confined medium.
We use anhydrous grade n-hexadecane (99+\%, Sigma-Aldrich) filtered through a 0.2 $\mu$m teflon membrane. A droplet
of the liquid (30--50 $\mu$L) is injected between the mica surfaces, a beaker containing phosphorus pentoxide is placed
 inside the vessel which is then sealed and left for thermal equilibration for 12h before starting experiments.
 
 \subsection{Layering under confinement}
 
\begin{figure}[htbp]
$$
\includegraphics[width=7cm]{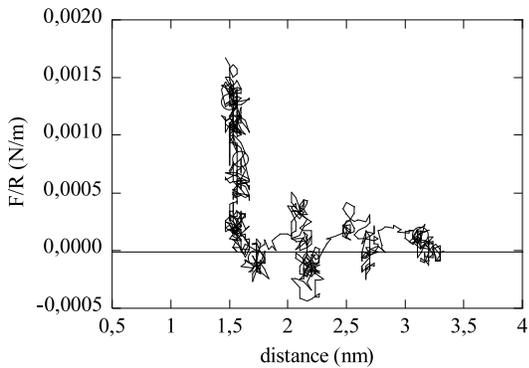}
$$
\caption{Normalized force (normal load divided by radius of curvature) as a function of film thickness, for hexadecane confined between mica surfaces, at 24$^{\circ}$C. The curve
corresponds to the approach phase only. Data acquisition rate of 13 Hz.}
\label{fig:fig2}
\end{figure}

 We first report the force-distance curve obtained when the surfaces are approached by driving the free end of the normal spring at a constant speed of 2\AA.s$^{-1}$. 
 Figure \ref{fig:fig2} clearly shows that the alkane exhibits layering under confinement, as seen from the 0.4--0.5 nm jumps in the intersurface distance, in
 agreement with previous studies of the same liquid \cite{HKC1,perez}. 
 
\begin{figure}[htbp]
$$
\includegraphics[width=7cm]{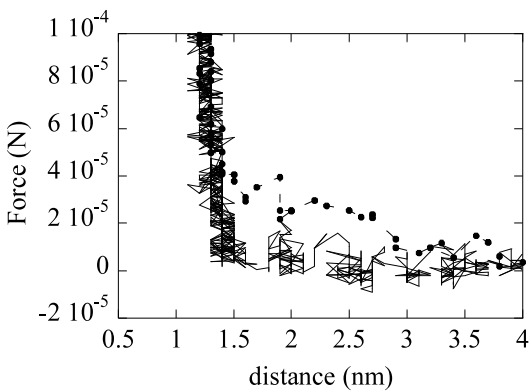}
$$
\caption{Normal load as a function of hexadecane thickness for two approach velocities: 0.2 nm.s$^{-1}$ (solid line) and 5 nm.s$^{-1}$ (dashed line with symbols).}
\label{fig:fig3}
\end{figure}

The present experimental setup allows for straightforward studies of dynamical effects which manifest when the fluid is confined at high enough velocities, and this without the need 
for post-analysis of recorded FECO patterns. This is illustrated on figure \ref{fig:fig3} where we have plotted two approach curves obtained at 0.2 and 5 nm.s$^{-1}$. It is seen on 
Fig. \ref{fig:fig3} that for thicknesses ranging from 1 to 4 nm, high velocity confinement leads to much less molecular layering and to an upward shift of the force-distance profile, caracterizing
the out-of-equilibrium response of the fluid. 

\subsection{Large strain shear experiments}

We now illustrate the shear capabilities of the instrument. 

\begin{figure}[htbp]
$$
\includegraphics[width=7cm]{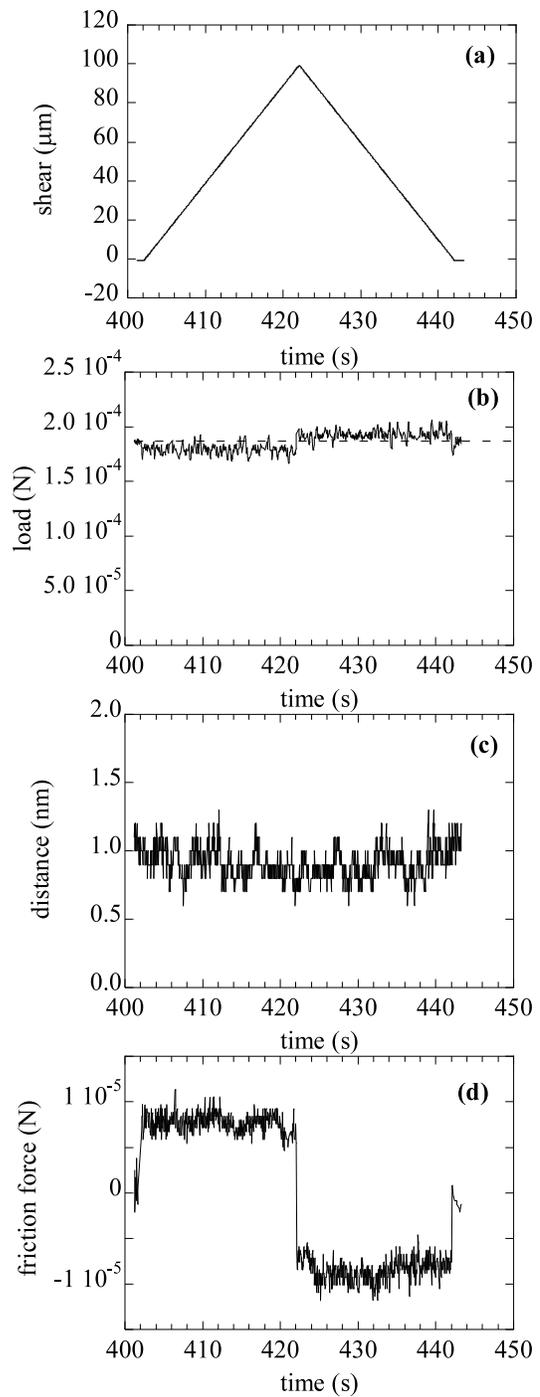}
$$
\caption{Friction experiment at controlled normal load over a sliding distance of 100 $\mu$m. (a) shear displacement. (b) normal load (the horizontal dashed line indicate the setpoint). 
(c) distance (the apparent increase/decrease of distance on a 20 s time scale actually corresponds to the point of closest distance  moving slightly off-axis of the microscope objective during shear). 
(d) friction force.}
\label{fig:fig4}
\end{figure}

On figure \ref{fig:fig4}, we show an
experiment where a confined film of hexadecane is sheared at a speed of 5 $\mu$m.s$^{-1}$ over a total sliding distance of 100 $\mu$m. This experiment was performed with a
normal load setpoint of 185 $\mu$N and an acquisition rate of 25 Hz. Under such conditions, the normal load is found to settle at 178 $\mu$N during forward shear, and at 192 $\mu$N
during backward motion (Fig. \ref{fig:fig4}b). Increasing further the integral gain of the feedback to reduce the offset to the setpoint was found to destabilize the control loop. 
The film thickness is measured to be 9 \AA, and the noise amplitude of $\pm 2$ \AA~visible on Fig. \ref{fig:fig4}c is not affected by shear motion and corresponds to the measurement sensitivity at 
the chosen 
acquisition rate. Steady sliding is observed under a shear force of 8 $\mu$N (Fig. \ref{fig:fig4}d). 
In open-loop conditions, {\it i.e.} without normal load feedback, the non-parallelism of the surfaces ($5\times 10^{-3}$ rad for this experiment) would yield a load variation of about 15 $\mu$N/$\mu$m, 
resulting in a complete loss of confinement after 20 $\mu$m of forward shear.

\begin{figure}[htbp]
$$
\includegraphics[width=7cm]{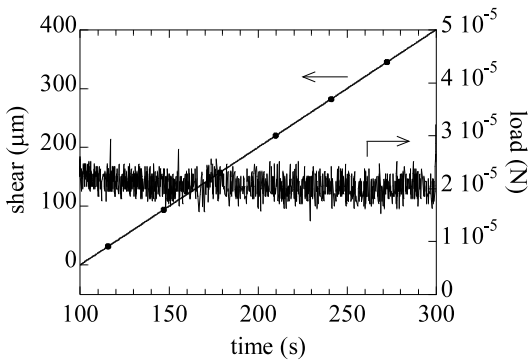}
$$
\caption{Shear experiment at controlled normal load over a sliding distance of 400 $\mu$m. Left scale: shear motion. Right scale: normal load.}
\label{fig:fig5}
\end{figure}

Larger shear amplitudes and lower load levels can readily be achieved. This is 
illustrated on figure \ref{fig:fig5}, where a load of 21$\pm$3 $\mu$N (setpoint 24 $\mu$N) is maintained over 400 $\mu$m at a sliding speed of
2 $\mu$m.s$^{-1}$.

 \begin{figure}[htbp]
$$
\includegraphics[width=7cm]{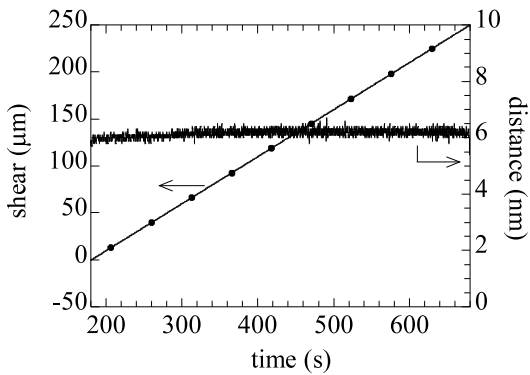}
$$
\caption{Shear experiment at controlled intersurface distance over a sliding distance of 250 $\mu$m. Left scale: shear motion. Right scale: distance.}
\label{fig:fig6}
\end{figure}

Experiments can also be performed under low confinement, by using the intersurface distance as the input signal of the
feedback control. On figure \ref{fig:fig6}, we show such an experiment where a film of hexadecane is sheared over 250
 $\mu$m at a velocity of 0.5 $\mu$m.s$^{-1}$, while maintaining its thickness at 60$\pm$1.5 \AA (distance setpoint 60 \AA). Under these 
 confinement conditions, normal and shear forces are below the instrument resolution.
 
The maximum acquisition rate of 27 Hz for simultaneous measurement of forces and distance limits the overall
performances of the digital feedback loop and the range of accessible sliding speed. Typically, fluctuations of 20--30 $\mu$N or 10--20 \AA~are observed on normal force or distance when using a sliding speed of 
20 $\mu$m.s$^{-1}$. No noticeable noise is induced by the feedback control when the sliding velocity stays below 2 
$\mu$m.s$^{-1}$.

\section{Conclusions}

We have built a surface force apparatus specially designed for shear experiments over large sliding distances ($>$100 $\mu$m). Feedback control is used in order to keep the normal load or the thickness of the 
confined medium constant during motion. The instrument exhibits exceptional performances in terms of force or distance stability, 
even in situations of weak confinement under small or zero 
applied load. We now plan to use this apparatus to investigate the shear behavior of glassy polymer thin films under
low pressure, and more generally of media presenting a complex molecular architecture, where large shear strains are needed in order to establish a steady-state regime.

\section{Aknowledgements}

We wish to thank Eric Perez, Philippe Richetti and Carlos Drummond for valuable discussions and for their advice during the development of the instrument.


\end{document}